\documentclass[hyper]{prop2015}
\usepackage{amsmath,amsfonts,amssymb,amsthm}
\usepackage{bbm}
\usepackage[all]{xy}
\usepackage{tikz}
\usetikzlibrary{arrows}
\usetikzlibrary{patterns}
\usetikzlibrary{decorations.markings}

\newcommand{\maps}{\colon}    
\newcommand{\R}{{\mathbbm R}}  
  
\newcommand{\Z}{{\mathbbm Z}}  
\newcommand{\N}{{\bf N}} 
\newcommand{\typeIIA}{\R^{9,1|{\bf 16} + \overline{\bf 16}}} 
\newcommand{\typeM}{\R^{10,1|{\bf 32}}}
\newcommand{\even}{{\rm even}}
\newcommand{\odd}{{\rm odd}}

\newcommand{\U}{{\rm U}}    
\newcommand{\SO}{{\rm SO}}

\newcommand{\Spin}{{\rm Spin}}

\newcommand{\g}{\mathfrak{g}}

\newcommand{\Cl}{{\rm C\ell}}
\newcommand{\Aut}{\mathrm{Aut}}

\newcommand{\tensor}{\otimes}

\newcommand{\define}[1]{{\bf \boldmath{#1}}}

\newtheorem{thm}{Theorem}

\newtheorem{prop}[thm]{Proposition}

\theoremstyle{definition}

\newcommand{\be}{\begin{equation}}
\newcommand{\ee}{\end{equation}}
\newcommand{\ba}{\begin{eqnarray}}
\newcommand{\ea}{\end{eqnarray}}
\newcommand{\ban}{\begin{eqnarray*}}
\newcommand{\ean}{\end{eqnarray*}}
\newcommand{\barr}{\begin{array}}
\newcommand{\earr}{\end{array}}

\category{Proceedings}
\subtitle{\href{http://www.maths.dur.ac.uk/lms/109/index.html}{LMS/EPSRC Durham Symposium on Higher Structures in M-Theory}}

\title{How Space-Times Emerge from the Superpoint}

\author[J.~Huerta]{John Huerta \inst{a,}\footnote{Corresponding author e-mail:~\href{mailto:jhuerta@math.tecnico.ulisboa.pt}{jhuerta@math.tecnico.ulisboa.pt}}}

\begin{acknowledgements}
 This work was partially supported by the FCT project {\em Higher Structures
and Applications}, with reference number PTDC/MAT-PUR/31089/2017.
\end{acknowledgements}

\shortauthors{J.~Huerta}

\address[1]{CAMGSD, Instituto Superior T\'ecnico, Av. Ravisco Pais, 1049-001 Lisboa, Portugal}

\begin{abstract}
  We describe how the super Minkowski space-times relevant to string theory and
  M-theory, complete with their Lorentz metrics and spin structures, emerge from a
  much more elementary object: the superpoint. In the sense of higher structures,
  this comes from treating the superpoint as an object in a flavor of rational
  homotopy theory, and repeatedly constructing central extensions. We will fit this
  story into the larger picture of the brane bouquet of Fiorenza--Sati--Schreiber:
  string theories and membrane theories emerge from super Minkowski space-times in
  precisely the same way as the super Minkowski space-times themselves emerge from the
  superpoint. This note is adapted from a talk I gave at the Durham symposium {\em
  Higher Structures in M-Theory}.
\end{abstract}
\shortabstract

\begin{document}

\maketitle

\section{Introduction}

The superpoint, denoted $\R^{0|1}$, is the space with a single odd coordinate
$\theta$. Because it is odd, its square vanishes, $\theta^2 = 0$, and a power series
expansion terminates at first order:
\begin{equation} f(\theta) = f(0) + f'(0) \theta . \end{equation}
For this reason, we regard $\theta$ as infinitesimal. Geometrically, the space
$\R^{0|1}$ consists of a single point with an infinitesimal neighborhood around it.

We can probe this straightforward space with the tools of homotopy theory. When we do
this, we discover something remarkable: all the super Minkowski spacetimes of
importance to string theory and M-theory, including their metrics and spin
structures, can be constructed as extensions of the superpoint. From these
space-times, using the brane bouquet of Fiorenza--Sati--Schreiber, we then find the
strings and branes themselves.

This note is a gentle introduction to these ideas. It is based on a talk I gave at
the Durham symposium {\em Higher Structures and M-theory} in August 2018, and that
talk was about work with Schreiber \cite{Huerta:2017utu}. Although our results
concern M-theory, a part of physics, our techniques are pure mathematics. Let us
nonetheless begin with the physical motivation.

\section{M-theory}

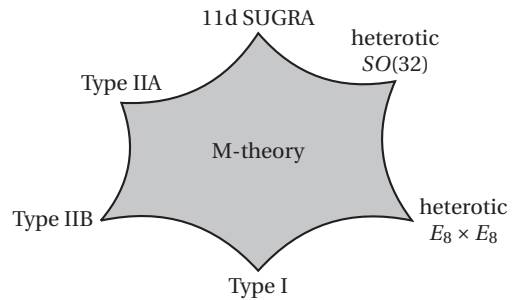
\begin{figure}[h]
\begin{center}
\begin{tikzpicture}[scale=.9,every node/.style={scale=.9}]
\node (11DSUGRA) [align=center] at (0,2) {11d SUGRA};
\node (TIIA) [align=center] at (-2,1) {Type IIA};
\node (TIIB) [align=center] at (-3,-1) {Type IIB};
\node (SO32) [align=center] at (2,1.5) {heterotic\\$SO(32)$};
\node (E8E8) [align=center] at (3,-1) {heterotic\\$E_8 \times E_8$};
\node (TI) [align=center] at (0,-2) {Type I};
\filldraw[color=gray!50,bend right] (TIIB.east) to (TIIA.south) to (11DSUGRA.south) to (SO32.south) to (E8E8.west) to (TI.north) to  (TIIB.east);
\draw[bend right,thick] (TIIB.east) to (TIIA.south) to (11DSUGRA.south) to (SO32.south) to (E8E8.west) to (TI.north) to  (TIIB.east);
\node (MTH) [align=center] at (0,0) {M-theory};
\end{tikzpicture}
\end{center}
\vspace{-.5cm}
\caption{Schematic for M-theory. Depicted are the different superstring theories and 11d supergravity as limit of an underlying theory---M-theory. }
\label{fig:schematic}
\end{figure}

In the 1990s, the string theory community realized they had to study objects of
dimension larger than 1, called \define{branes}. Witten christened this topic
\define{M-theory} \cite{Witten:1995ex}, where the M arguably stands for `membrane'
\cite{Duff:1999baa}. The idea of M-theory, not yet fully realized today, is that it should be
single physical theory having the five superstring theories in 10d as limits, and its
classical limit should be 11d supergravity. This idea is often pictured schematically
as in Figure \ref{fig:schematic}. 

\begin{figure}[h]
\begin{center}
\begin{tikzpicture}[scale=.9,every node/.style={scale=.9}]
\node (11DSUGRA) [align=center] at (0,2) {11d SUGRA};
\node (TIIA) [align=center] at (-2,1) {Type IIA};
\node (TIIB) [align=center] at (-3,-1) {Type IIB};
\node (SO32) [align=center] at (2,1.5) {heterotic\\$SO(32)$};
\node (E8E8) [align=center] at (3,-1) {heterotic\\$E_8 \times E_8$};
\node (TI) [align=center] at (0,-2) {Type I};
\filldraw[color=gray!50,bend right] (TIIB.east) to (TIIA.south) to (11DSUGRA.south) to (SO32.south) to (E8E8.west) to (TI.north) to  (TIIB.east);
\draw[bend right,thick] (TIIB.east) to (TIIA.south) to (11DSUGRA.south) to (SO32.south) to (E8E8.west) to (TI.north) to  (TIIB.east);
\draw[bend right,thick,->] (TIIA.east) to (-.4,1.6);
\node (MTH) [align=center] at (0,0) {M-theory};
\node (LIMIT) [align=center] at (-1,1.38) {$N$};
\end{tikzpicture}
\end{center}
\vspace{-.5cm}
  \caption{11d physics as a `large-$N$ limit' of type IIA superstring theory, where $N$ is the number of D$0$-branes; see Polchinski~\cite{Polchinski:1999br} for details.}  \label{fig:schematiclimit}
\end{figure}
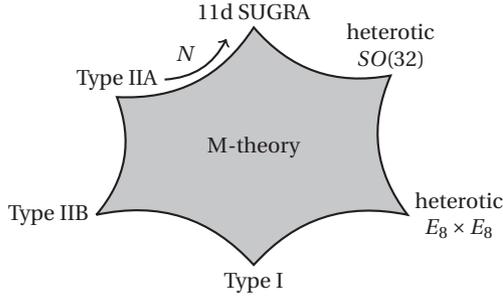

The idea that we can have an 11d theory and 10d theory on equal footing shows off how the concept of `dimension' is
flexible in M-theory. While M-theory is posited to have these various limits, in
practice we do not even know its degrees of freedom. What we glimpse of M-theory
comes in fact from taking certain limits of string theories. Most directly, we can
take a certain limit of 10d type IIA string theory to obtain 11d physics, as
pictured in Figure \ref{fig:schematiclimit}.

This suggests there must be some mathematical process that turns the 10d space-time of
string theory into the 11d space-time of M-theory:
\begin{equation} M^{9,1} \rightsquigarrow N^{10,1} . \end{equation}
Infinitesimally, at tangent spaces, this process turns 10d Minkowski space-time into
11d:
\begin{equation} \R^{9,1} \rightsquigarrow \R^{10,1} . \end{equation}
But string theory and M-theory are both supersymmetric, so really this should be
between super Minkowski space-times:
\begin{equation} \typeIIA \rightsquigarrow \typeM . \end{equation}
In this last case, there is a natural mathematical choice for this process: it is
called central extension. To understand it, we need to understand super Minkowski
space-time.

\section{Super Minkowski space-time}

Super Minkowski space-time $\R^{d-1,1|\N}$ is the supersymmetric version of Minkowski
space-time, $\R^{d-1,1}$, which itself is just $\R^d$ with the indefinite metric
$\eta(u,v) = -u^0 v^0 + u^1 v^1 + \cdots + u^{d-1} v^{d-1}$. Mathematically,
$\R^{d-1,1|\N}$ is a \define{super Lie algebra}. This is a Lie algebra in super
vector spaces, so $\R^{d-1,1|\N}$ has an underlying super vector space:
\begin{equation} \R^{d-1,1|\N}_\even = \R^{d-1,1}, \quad \R^{d-1,1|\N}_\odd = \N . \end{equation}
The even part is ordinary Minkowski space-time $\R^{d-1,1}$. The odd part, $\N$, is a
new ingredient: it is a spinor representation.

On this super vector space, we have a bracket:
\begin{equation} [-,-] \maps \R^{d-1,1|\N} \otimes \R^{d-1,1|\N} \to \R^{d-1,1|\N} . \end{equation}
This bracket satisfies the axioms of a Lie algebra, up to some signs. The bracket,
and indeed the whole theory of super Minkowski space-time, is governed by the
representation theory of the spin group, $\Spin(d-1,1)$. This Lie group is the double
cover of the connected Lorentz group, $\SO_0(d-1,1)$.

The spin group $\Spin(d-1,1)$ has more representations than $\SO_0(d-1,1)$. In
particular, it has `spinor representations': there is a well-known inclusion
$\Spin(d-1,1) \hookrightarrow \Cl(d-1,1)$ of the spin group into the Clifford
algebra. In fact, the Clifford algebra is $\Z_2$-graded, and the spin group lands in
the even subalgebra $\Spin(d-1,1) \hookrightarrow \Cl_0(d-1,1)$. Modules of
$\Cl_0(d-1,1)$ thus become representations of $\Spin(d-1,1)$. These are
\define{spinor representations}. Our $\N$ is one of these.

In detail:
\smallskip
\begin{enumerate}[i)]
\item $\Spin(d-1,1)$ acts on $\R^{d-1,1}$ by linear transformations preserving the metric;
\item $\N$ is a real spinor representation of $\Spin(d-1,1)$;
\item The bracket $[-,-] \maps \R^{d-1,1|\N} \otimes \R^{d-1,1|\N} \to \R^{d-1,1|\N}$
  is equivariant with respect to the action of $\Spin(d-1,1)$;
\item In fact, the only nonzero part of the bracket is the spinor-to-vector pairing:
  \begin{equation} [-,-] \maps \N \otimes \N \to \R^{d-1,1} . \end{equation}
\end{enumerate}
\smallskip
Physicists write the bracket as $[Q_\alpha, Q_\beta] = -2\Gamma^\mu_{\alpha \beta}
P_\mu$, where $\{Q_\alpha\}$ is a basis for $\N$, $\{P_\mu\}$ is a basis for
$\R^{d-1,1}$, and $\Gamma$ is the gamma matrix for our spinor representation. When
$\N$ is irreducible, the bracket is the unique equivariant map up to
rescaling. Otherwise it involves a choice.

Because super Minkowski space-times are super Lie algebras, we can consider their
central extensions. This will provide our sought after mathematical process going
from 10d to 11d:
\begin{equation} \typeIIA \rightsquigarrow \typeM . \end{equation}
Here's how it works: for any super Lie algebra $\g$, a \define{central extension} is
a short exact sequence of super Lie algebras:
\begin{equation} 0 \longrightarrow \R \longrightarrow \tilde{\g} \longrightarrow \g \longrightarrow
  0 \end{equation}
such that $\R$ lands in the center of $\tilde{\g}$. Mathematically, central
extensions are classified by super Lie algebra cohomology, and this allows us to
describe them very concretely. Specifically, \define{2-cocycle} on $\g$ is
skew-symmetric map:
\begin{equation} \omega \maps \Lambda^2 \g \to \R \end{equation}
satisfying the \define{cocycle condition}:
\begin{equation} \omega([X,Y],Z) \pm \omega([Y,Z],X) \pm \omega([Z,X],Y) = 0 \end{equation}
where the signs depend on whether $X, Y, Z$ are even or odd elements of $\g$. Given
one of these 2-cocycles, we can define a central extension $\g_\omega$ to be the
super Lie algebra obtained from $\g$ by including one extra generator $c$, even and
central:
\begin{equation} \g_\omega = \g \oplus \R c \end{equation}
and modifying the bracket with the 2-cocycle:
\begin{equation} [X,Y]_\omega = [X,Y] + \omega(X,Y) c \mbox{ for } X, Y \in \g. \end{equation}
Since $c$ is central, this defines the bracket on all of $\g_\omega$. We thus get a
central extension:
\begin{equation} 0 \longrightarrow \R \longrightarrow \g_\omega \longrightarrow \g \longrightarrow
  0 \end{equation}
where $\R$ is included as $\R c$, and $\g_\omega \longrightarrow \g$ is the map setting $c$ to
zero. Every central extension is isomorphic to one of this form: thus, 2-cocycles
give us a central extensions, and vice versa.  In what follows, we will often denote
a central extension by the homomorphism $\g_\omega \longrightarrow \g$ which sets $c$
to zero.

The cocycle condition may look mysterious, but it is exactly what we need to
guarantee that the Lie bracket on $\g_\omega$ satisfies the Jacobi identity. It also
has a beautiful geometric interpretation: if $G$ is a super Lie group with super Lie
algebra $\g$, then $\omega$ defines a 2-form on $G$ by left-translation. The cocycle
condition then holds if and only if this form is {\em closed}: ${\rm d} \omega = 0$.

Thus, all we need to extend from 10d to 11d is a 2-cocycle. Here is one, written as a
2-form:
\begin{equation} \omega = {\rm d} \theta^\alpha \wedge \Gamma^{01 \cdots 9}_{\alpha \beta} {\rm d} \theta^\beta
  , \end{equation}
where $\Gamma^{01 \cdots 9} = \Gamma^0 \Gamma^1 \cdots \Gamma^9$ is the product of
all the gamma matrices, and $\{\theta^\alpha\}$ are odd coordinates on
$\typeIIA$. This 2-form is indeed left-invariant on the super Lie group corresponding
to $\typeIIA$, and $d \omega = 0$ by the naive calculation. Hence, $\omega$ is in
fact a 2-cocycle. Finally, centrally extending $\typeIIA$ by this 2-cocycle does
indeed yield $\typeM$, by some Clifford algebraic yoga. We thus have the central
extension:
\begin{equation} \typeM \longrightarrow \typeIIA . \end{equation}

This example raises a few questions. First, why should we use the 2-cocycle $\omega$?
There could be others on\linebreak $\typeIIA$. What singles out $\omega$? The answer is an
invariance condition: $\omega$ is invariant under the action of $\Spin(9,1)$. Next,
can we account for more dimensions in space-time by central extension? Taking this to
extremes, can we realize all the space-times we care about by centrally extending the
superpoint, $\R^{0|1}$? Indeed we can; this is our main result.

To begin, let us define the superpoint more precisely: the \define{superpoint}
$\R^{0|1}$ is the super vector space with vanishing even degree and $\R$ in odd
degree:
\begin{equation} \R^{0|1}_\even = 0 , \quad \R^{0|1}_\odd = \R . \end{equation}
It is crucial to note that $\R^{0|1}$ has no Lie bracket, no metric, and no spin
structure. We will discover all of these by central extension.

Despite this lack of structure, the superpoint has a 2-cocycle:
\begin{equation} {\rm d}\theta \wedge {\rm d} \theta . \end{equation}
This is nonzero precisely because $\theta$ is odd. Centrally extending by this
cocycle, we get $\R^{1|1}$, the worldline of the super particle:
\begin{equation} \R^{1|1} \longrightarrow \R^{0|1} . \end{equation}
That is already something, but we can go a lot further.

\section{The dimensional ladder}

Let us play a game with two moves, starting with the superpoint:
\smallskip
\begin{enumerate}[i)]
  \item extend by all nontrivial 2-cocycles subject to a suitable invariance condition;
  \item if no 2-cocycles are available, double the number of odd dimensions. 
\end{enumerate}

\smallskip
We need to spell out that invariance condition. We want to say that our 2-cocycles
need to be Lorentz-invariant, or more precisely, invariant under the spin group. But
we cannot, because there is no metric as yet. Fortunately, the symmetries of the
metric turn out to be encoded in the Lie bracket:
\begin{prop}[H.--Schreiber, folklore]
  For a super Min\-kowski space-time $\R^{d-1,1|\N}$, its connected automorphism group
  is, up to cover:
  \begin{equation} \Aut_0(\R^{d-1,1|\N}) \cong \R^+ \times \Spin(d-1,1) \times \mbox{R-group} \end{equation}
  where the R-group acts trivially on $\R^{d-1,1}$.
\end{prop}

The R-group is known, in the physics literature, as the R-symmetries of
$\R^{d-1,1|\N}$. We can generalize this to any super Lie algebra: an
\define{R-symmetry} of $\g$ is an automorphism that acts trivially on
$\g_\even$. This theorem was probably folklore among physicists, but not finding a
proof in the literature, we went ahead and proved it. Its significance is that it
allows us to get our hands on the $\Spin(d-1,1)$ symmetries, in the terms of the
super Lie algebra structure alone, without mentioning the metric. Specifically: a
2-cocycle on a super Lie algebra $\g$ is called \define{invariant} if it is invariant
under the quotient of $\Aut_0(\g)$ by rescalings and R-symmetries. With this
definition, when $\g$ is a super Minkowski space-time, a 2-cocycle is invariant
precisely when it is invariant under $\Spin(d-1,1)$.

Let us begin. First, we will double the number of odd dimensions of $\R^{0|1}$,
yielding $\R^{0|2}$.  We will write this operation as follows:
\begin{equation} \kern-6pt\xymatrix{ \mathbbm{R}^{0 \vert 2} \ar@{<-}@<-3pt>[r] \ar@{<-}@<+3pt>[r] &
  \mathbbm{R}^{0\vert 1} } \end{equation}
Now, $\R^{0|2}$ has two odd generators, $\theta_1$ and $\theta_2$, and there are
three 2-cocycles:
\begin{equation} {\rm d} \theta_1 \wedge {\rm d} \theta_1, \quad {\rm d}\theta_1 \wedge {\rm d}\theta_2, \quad {\rm d}\theta_2
  \wedge {\rm d}\theta_2 . \end{equation}
Because $\R^{0|2}$ has no even part, any automorphism must be an R-symmetry. Hence,
all of these 2-cocycles are invariant under the maximal subgroup containing no
nontrivial R-symmetries. Extending by all three we get:
\begin{equation} \R^{3|2} \longrightarrow \R^{0|2} . \end{equation}
At this point, something remarkable happens: a metric appears,
\begin{equation} \Aut_0(\R^{3|2}) = \R^+ \times \Spin(2,1) . \end{equation}
We did not put it in, but by looking at the automorphisms of the algebra, the three
even generators in $\R^{3|2}$ transform under $\Spin(2,1)$ as vectors, and the two
odd generators as spinors.

Thanks to this metric, we can look for $\Spin(2,1)$-invariant 2-cocycles on
$\R^{2,1|2}$. There are none, because the only $\Spin(2,1)$-invariant map:
\begin{equation} \mathbf{2} \tensor \mathbf{2} \to \R \end{equation}
is antisymmetric.

Since we are out of 2-cocycles, let us double the number of odd dimensions again:
\begin{equation} \kern-6pt\xymatrix{ \mathbbm{R}^{2,1 \vert \mathbf{2} + \mathbf{2}} \ar@{<-}@<-3pt>[r]
  \ar@{<-}@<+3pt>[r] & \mathbbm{R}^{2,1 \vert \mathbf{2}} }\end{equation}
There is precisely one $\Spin(2,1)$-invariant 2-cocycle, and extending by this gives:
\begin{equation} \R^{3,1|\mathbf{4}} \longrightarrow \R^{2,1|\mathbf{2} + \mathbf{2}} \end{equation}
Again, the metric is not a choice:
\begin{equation} \Aut_0(\R^{3,1|\mathbf{4}}) = \R^+ \times \Spin(3,1) \times \U(1) .\end{equation}
Here, $\U(1)$ is the R-group. There are no further $\Spin(3,1)$-invariant 2-cocycles.

We can keep going in exactly this way, up to dimension 11. Two notable phenomena
occur. First, we sometimes encounter several 2-cocycles after doubling the number of
spinors: in dimensions 4, there are two 2-cocycles, so we jump directly to dimension
6. In dimension 6, after doubling, there are four 2-cocycles, so we jump directly to
dimension 10. Moreover, in dimensions 6 and 10, there are two distinct spinor
representations, so there are two ways to double, a type IIA, where we include both
kinds of spinor, and a type IIB, where we just include one kind. In summary, we have
the following collection of doublings and central extensions that we display in
Theorem \ref{thm:superpoint}.

\begin{thm}[H.--Schreiber] \label{thm:superpoint}
 \begin{equation} \hspace{0.5cm}
   \vspace{-0.5cm}
  \xymatrix@R=18pt@C=16pt{
    & \mathbbm{R}^{10,1\vert \mathbf{32}}
      \ar[ddr]
    &&&
    \\
    & 
    &&
    &
    \\
    \mathbbm{R}^{9,1 \vert \mathbf{16} + {\mathbf{16}}}
    \ar@{<-}@<-3pt>[r]
    \ar@{<-}@<+3pt>[r]
    & \mathbbm{R}^{9,1\vert \mathbf{16}} \ar[dr]
    \ar@<-3pt>[r]
    \ar@<+3pt>[r]
    &
    \mathbbm{R}^{9,1\vert \mathbf{16} + \overline{\mathbf{16}}}
    \\
    \mathbbm{R}^{5,1 \vert \mathbf{8} + \mathbf{8}}
    \ar@{<-}@<-3pt>[r]
    \ar@{<-}@<+3pt>[r]
    &
    \mathbbm{R}^{5,1\vert \mathbf{8}}
    \ar[dl]
    \ar@{->}@<-3pt>[r]
    \ar@{->}@<+3pt>[r]
    &
    \mathbbm{R}^{5,1 \vert \mathbf{8} + \overline{\mathbf{8}}}
    \\
    \mathbbm{R}^{3,1\vert \mathbf{4}+ \mathbf{4}}
    \ar@{<-}@<-3pt>[r]
    \ar@{<-}@<+3pt>[r]
    &
    \mathbbm{R}^{3,1\vert \mathbf{4}}
    \ar[dl]
    \\
    \mathbbm{R}^{2,1 \vert \mathbf{2} + \mathbf{2} }
    \ar@{<-}@<-3pt>[r]
    \ar@{<-}@<+3pt>[r]
    &
    \mathbbm{R}^{2,1 \vert \mathbf{2}}
    \ar[dl]
    \\
    \mathbbm{R}^{0 \vert \mathbf{1}+ \mathbf{1}}
    \ar@{<-}@<-3pt>[r]
    \ar@{<-}@<+3pt>[r]
    &
    \mathbbm{R}^{0\vert \mathbf{1}}
  }
\end{equation}
\end{thm}

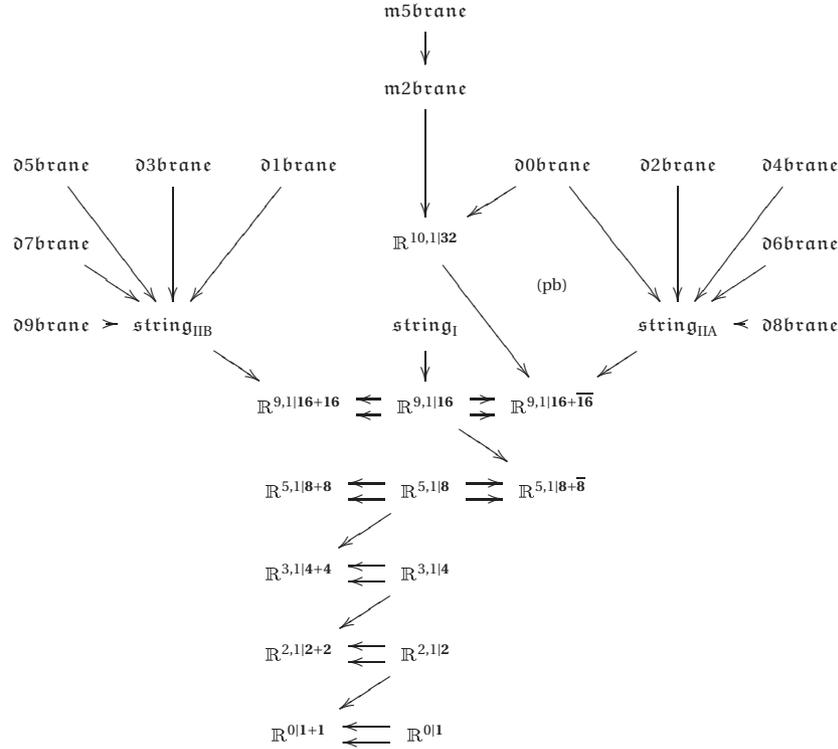
\begin{figure*}[h]
  \begin{equation*} \hspace{1cm}
    \def\objectstyle{\scriptstyle}
    \def\labelstyle{\scriptstyle}
    \kern1cm\xymatrix@C=4pt@R=12pt{
    &
    &&&& \mathfrak{m}5\mathfrak{brane}
     \ar[d]
    \\
    &
    &&
     && \mathfrak{m}2\mathfrak{brane}
    \ar[dd]
    &&
    \\
    &
    &
    \mathfrak{d}5\mathfrak{brane}
    \ar[ddr]
    &
    \mathfrak{d}3\mathfrak{brane}
    \ar[dd]
    &
    \mathfrak{d}1\mathfrak{brane}
    \ar[ddl]
    &
    & \mathfrak{d}0\mathfrak{brane}
    \ar@{}[ddd]|{\mbox{\tiny (pb)}}
    \ar[ddr]
    \ar[dl]
    &
    \mathfrak{d}2\mathfrak{brane}
    \ar[dd]
    &
    \mathfrak{d}4\mathfrak{brane}
    \ar[ddl]
    \\
    &
    &
    \mathfrak{d}7\mathfrak{brane}
    \ar[dr]
    &
    &
    & \mathbbm{R}^{10,1\vert \mathbf{32}}
      \ar[ddr]
    &&&
    \mathfrak{d}6\mathfrak{brane}
    \ar[dl]
    \\
    &
    &
    \mathfrak{d}9\mathfrak{brane}
    \ar[r]
    &
    \mathfrak{string}_{\mathrm{IIB}}
    \ar[dr]
    &
    & \mathfrak{string}_{\mathrm{I}}
      \ar[d]
    &&
    \mathfrak{string}_{\mathrm{IIA}}
    \ar[dl]
    &
    \mathfrak{d}8\mathfrak{brane}
    \ar[l]
    \\
    &
    &
    &
    &
    \mathbbm{R}^{9,1 \vert \mathbf{16} + {\mathbf{16}}}
    \ar@{<-}@<-3pt>[r]
    \ar@{<-}@<+3pt>[r]
    & \mathbbm{R}^{9,1\vert \mathbf{16}} \ar[dr]
    \ar@<-3pt>[r]
    \ar@<+3pt>[r]
    &
    \mathbbm{R}^{9,1\vert \mathbf{16} + \overline{\mathbf{16}}}
    \\
    &
    &
    &
    &
    \mathbbm{R}^{5,1 \vert \mathbf{8} + \mathbf{8}}
    \ar@{<-}@<-3pt>[r]
    \ar@{<-}@<+3pt>[r]
    &
    \mathbbm{R}^{5,1\vert \mathbf{8}}
    \ar[dl]
    \ar@{->}@<-3pt>[r]
    \ar@{->}@<+3pt>[r]
    &
    \mathbbm{R}^{5,1 \vert \mathbf{8} + \overline{\mathbf{8}}}
    \\
    &
    &
    &
    &
    \mathbbm{R}^{3,1\vert \mathbf{4}+ \mathbf{4}}
    \ar@{<-}@<-3pt>[r]
    \ar@{<-}@<+3pt>[r]
    &
    \mathbbm{R}^{3,1\vert \mathbf{4}}
    \ar[dl]
    \\
    &
    &
    &
    &
    \mathbbm{R}^{2,1 \vert \mathbf{2} + \mathbf{2} }
    \ar@{<-}@<-3pt>[r]
    \ar@{<-}@<+3pt>[r]
    &
    \mathbbm{R}^{2,1 \vert \mathbf{2}}
    \ar[dl]
    \\
    &
    &
    &
    &
    \mathbbm{R}^{0 \vert \mathbf{1}+ \mathbf{1}}
    \ar@{<-}@<-3pt>[r]
    \ar@{<-}@<+3pt>[r]
    &
    \mathbbm{R}^{0\vert \mathbf{1}}
} \end{equation*}

\caption{The brane bouquet}
\label{fig:bouquet}
\end{figure*}

\section{The brane bouquet}

In the last section, we saw what we could do with invariant 2-cocycles. Modulo a
suitable equivalence relation, 2-cocycles form a group, $H^2(\g)$, the second
cohomology group of the super Lie algebra $\g$. There are cohomology groups in higher
degree, $H^p(\g)$. Can we fit these into our story? What is the significance of
$p$-cocycles for $p \geq 3$?

There are two remarkable answers to this question, one coming from physics, and the
other from mathematics:
\begin{itemize}
\smallskip
\item[] {\bf Physics}: Invariant $(p+2)$-cocycles on $\R^{d-1,1|\N}$ correspond to
  Green--Schwarz $p$-branes \cite{DeAzcarraga:1989vh}.
  \smallskip
\item[] {\bf Mathematics}: Central extensions by $(p+2)$-cocycles on the super Lie algebra
  $\g$ yield `super $L_\infty$-algebras' \cite{Baez:2003aa}.
\end{itemize}
\smallskip
A \define{super $L_\infty$-algebra} $\g$ is like a Lie algebra, defined on a chain
complex of super vector spaces:
\begin{equation} \g_0 \stackrel{\partial}{\longleftarrow} \g_1 \stackrel{\partial}{\longleftarrow}
  \cdots \stackrel{\partial}{\longleftarrow} \g_n \stackrel{\partial}{\longleftarrow}
  \cdots \end{equation}
But the Jacobi identity \emph{does not hold}:
\begin{equation} [[X,Y],Z] \pm [[Y,Z],X] \pm [[Z,X],Y] \neq 0 . \end{equation}
Instead, it holds up to \emph{coherent homotopy}: we get infinitely many identities
like this:
\begin{equation} \begin{array}{c} [[X,Y],Z] \pm [[Y,Z],X] \pm [[Z,X],Y]=\ \\ = \partial [X,Y,Z] + [\partial(X \wedge Y
  \wedge Z)] \, . \end{array}  \end{equation}
  This says the Jacobi identity holds up to a chain homotopy, given by a trilinear
bracket:
\begin{equation} [-.-,-] \maps \g \otimes \g \otimes \g \to \g , \end{equation}
satisfying its own Jacobi-like identity up to a 4-linear bracket\ldots and so on,
forever.

The key insight of the brane bouquet due to Fiorenza--Sati--Schreiber \cite{Fiorenza:2013nha} is
that we can combine these two strands, one from physics and one from mathematics: we
can centrally extend by the higher degree cocycles classifying the Green--Schwarz
$p$-branes to obtain super $L_\infty$-algebras. Then we can look for additional
invariant cocycles on those $L_\infty$-algebras. Lo and behold, these new cocycles
turn out to correspond to additional branes, also very important in physics: D-branes
and the M5-branes. We can then centrally extend by these cocycles, and continue our
hunt for invariant cocycles, which should correspond to new branes.

Thus, by including higher degree cocycles, we get the brane bouquet, growing out of the superpoint as shown in Figure \ref{fig:bouquet}. There, we have named the super $L_\infty$-algebras after the physical objects to which their cocycles correspond.

In this note, I have recounted what we know so far. But I have not claimed to be
exhaustive: there may be more cocycles, and thus more extensions, waiting to be
found. A full computation of the brane bouquet has not been done. There may be many
more surprises waiting for us inside the humble superpoint.

\bibliography{allbibtex}
\bibliographystyle{prop2015}

\end{document}